\documentclass[twocolumn,showpacs,preprintnumbers,amsmath,amssymb]{revtex4}

\usepackage{epsfig}
\usepackage{bm}

\begin{document}

\preprint{Submitted to {\it Physical Review Letters}}

\title{Two-scale structure of the electron dissipation region during collisionless magnetic reconnection}

\author{M. A. Shay} \email{shay@udel.edu}
\homepage{http://www.physics.udel.edu/~shay}
\affiliation{Department of Physics \& Astronomy, 217 Sharp Lab,
  University of Delaware, Newark, DE 19716}
\author{J. F. Drake, M. Swisdak}
\affiliation{
University of Maryland, College Park, MD, 20742
}

\date{\today}

\begin{abstract}
Particle in cell (PIC) simulations of collisionless magnetic
reconnection are presented that demonstrate that the electron
dissipation region develops a distinct two-scale structure along the
outflow direction. The length of the electron current layer is found
to decrease with decreasing electron mass, approaching the ion
inertial length for a proton-electron plasma. A surprise, however, is
that the electrons form a high-velocity outflow jet that remains
decoupled from the magnetic field and extends large distances
downstream from the x-line. The rate of reconnection remains fast in
very large systems, independent of boundary conditions and the mass of
electrons.
\end{abstract}

\pacs{Valid PACS appear here}

\maketitle



Magnetic reconnection drives the release of magnetic energy in
explosive events such as disruptions in laboratory experiments,
magnetic substorms in the Earth's magnetosphere and flares in the
solar corona. Reconnection in these events is typically collisionless
because reconnection electric fields exceed the Dreicer runaway field.
Since magnetic field lines reconnect in a boundary layer, the
``dissipation region'', whose structure may limit the rate of release
of energy, understanding the structure of this boundary layer and its
impact on reconnection is critical to understanding the observations.
Because of their ability to carry large currents the dynamics of
electrons continues to be a topic of interest. Early simulations of
reconnection suggested that the rate of reconnection was not sensitive
to electron dynamics \cite{Hesse99,Shay98b} and this insensitivity was
attributed to the coupling to whistler dynamics at the small spatial
scales of the dissipation region \cite{Birn01,Rogers01}. The results
of more recent kinetic PIC simulations have called into question these
results by suggesting that the electron current layer stretches along
the outflow direction and the rate of reconnection
drops\cite{Daughton06,Fujimoto06}. The fast rates of reconnection
obtained from earlier simulations\cite{Shay99,Hesse99,Birn01} were
attributed to the influence of periodicity\cite{Daughton06}.

We present particle-in-cell (PIC) simulations with various electron
masses and computational domain sizes and an analytic model that
demonstrate that collisionless reconnection remains fast even in very
large collisionless systems. The reconnection rate stabilizes before
the periodicity of the boundary conditions can impact the dynamics.
The electron current layer develops a distinct two-scale structure
along the outflow direction that had not been identified in earlier
simulations. The out-of-plane electron current driven by the
reconnection electric field has a length that decreases with the
electron mass, scaling as $(m_e/m_i)^{3/8}$, which extrapolates to
about an ion inertial length $d_i=c/\omega_{pi}$ for the
electron-proton mass ratio. The surprise is that a jet of outflowing
electrons with velocity close to the electron Alfven speed $c_{Ae}$
extends up to several $10$'s of $d_i$ from the x-line. Remarkably, the
electrons are able to jet across the magnetic field over such enormous
distances because momentum transport transverse to the jet effectively
``blocks'' the flow of the out-of-plane current in this region. The
momentum transport causing this ``current blocking'' effect has the
same source (the off diagonal pressure tensor\cite{Hesse99}), but is
much stronger than that which balances the reconnection electric field
at the x-line.

Our simulations are performed with the particle-in-cell code p3d
\cite{Shay01,Zeiler02}. The results are presented in normalized units:
the magnetic field to the asymptotic value of the reversed field, the
density to the value at the center of the current sheet minus the
uniform background density, velocities to the Alfv\'en speed $v_A$,
lengths to the ion inertial length $d_i$, times to the inverse ion
cyclotron frequency $\Omega_{ci}^{-1}$, and temperatures to $m_i
v_A^2$. We consider a system periodic in the $x-y$ plane where flow
into and away from the x-line are parallel to $\mathbf{\hat{y}}$ and
$\mathbf{\hat{x}}$, respectively.  The reconnection electric field is
parallel to $\mathbf{\hat{z}}$. The initial equilibrium consists of
two Harris current sheets superimposed on a ambient population of
uniform density.  The reconnection magnetic field is given by
$B_x=\tanh[(y-L_y/4)/w_0]- \tanh[(y-3L_y/4)/w_0]-1$, where $w_0$ and
$L_y$ are the half-width of the initial current sheets and the box
size in the $\mathbf{\hat{y}}$ direction.  The electron and ion
temperatures, $T_e = 1/12$ and $T_i = 5/12$, are initially uniform.
The initial density profile is the usual Harris form plus a uniform
background of $0.2$. The simulations presented here are
two-dimensional,{\it i.e.}, $\partial/\partial z = 0$.  Reconnection
is initiated with a small initial magnetic perturbation that produces
a single magnetic island on each current layer.

\begin{figure}
  \epsfig{file=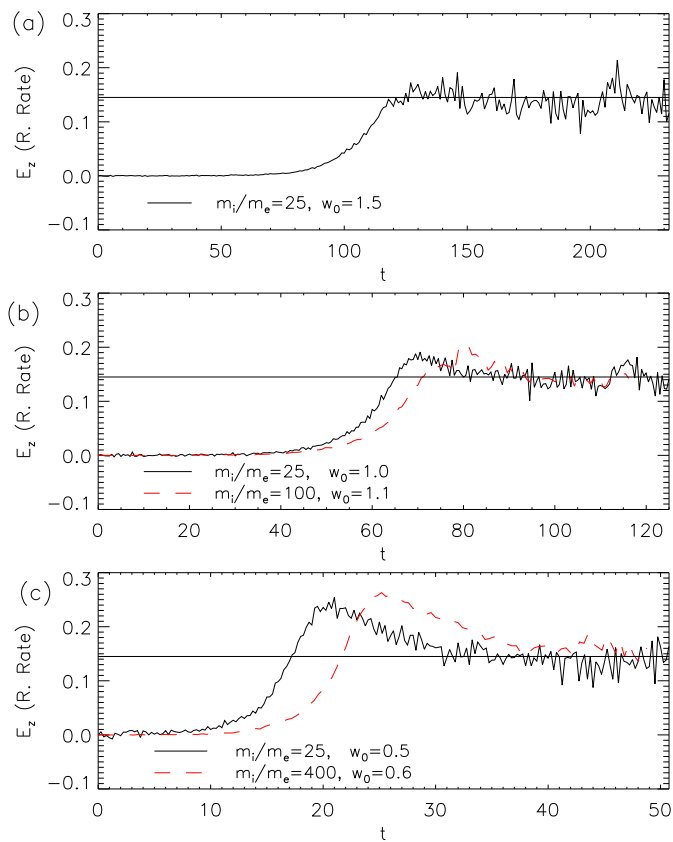,width=9cm}
  \caption{(color online). Reconnection electric field versus time: (a) $204.8
    \times 102.4,$ (b) $102.4 \times 51.2,$ (c) $51.2 \times 25.6.$ 
    $w_0$ is the initial current sheet width.}
  \label{rrate}
\end{figure}

We have explored the dependence of the rate of reconnection on the
system size in a series of simulations with three different system
sizes and three different mass ratios. For $m_i/m_e=25,$ the grid
scale $\Delta=0.05$ and the speed of light $c = 15.$ For
$m_i/m_e=100,$ $\Delta=0.025$ and $c=20.$ For $m_i/m_e=400,$
$\Delta=0.0125$ and $c=40.$ The reconnection rate versus time is
plotted for our simulations in Fig.~\ref{rrate}. The reconnection rate
is determined by taking the time derivative of the total magnetic flux
between the x-line and the center of the magnetic island.  The rate
increases with time, undergoes a modest overshoot that is more
pronounced in the smaller domains, and approaches a quasi-steady rate
of around $0.14$, independent of the domain size. Earlier suggestions
\cite{Daughton06} that reconnection rates would plunge until elongated
current layers spawned secondary magnetic islands are not borne out in
these simulations.  The rates of reconnection approach constant values
even in the absence of secondary islands, which for anti-parallel
reconnection typically only occur transiently due to initial
conditions\cite{Drake06}. Even these transient islands can be largely
eliminated by a suitable choice of the initial current layer width
$w_0$ (a larger value of $w_0$ is required for the larger domains).

A critical issue is whether the periodicity in the $x$ direction can
influence the rate of reconnection \cite{Daughton06}. In each of the
simulations we have identified the time at which the ion outflows from
the x-line meet at the center of the magnetic island. This occurs at
$t \approx 155$ for the largest simulation shown in Fig.~\ref{rrate}a.
The plasma at the x-line can not be affected by the downstream
conditions until $t \approx 255$, when a pressure perturbation can
propagate back upstream to the x-line at the magnetosonic speed. This
is well after the end of the simulation. The electrons are ejected
from the x-line at a velocity of around $c_{Ae}\gg c_A$ and therefore
might be able to follow field lines back to the x-line. During the
traversal time $\delta t=L_x/c_{Ae}$, the amount of reconnected flux
is $v_{in}B_0L/c_{Ae}$, where $v_{in}$ is the inflow velocity into the
x-line. Using the conservation of the canonical momentum in the
z-direction, the condition that an electron with a velocity $c_{Ae}$
can not cross this flux to access the x-line reduces to
$L>d_i(c_{A}/v_{in})\sim 7d_i$, which is easily satisfied for the
simulations in Fig.~\ref{rrate}. The fact that the reconnection rates
for all of the simulation domains in Fig.~\ref{rrate} are essentially
identical further supports this conclusion.

Also shown in Fig.~\ref{rrate} in the dashed lines are the rates of
reconnection for $m_i/m_e=100$ in (b) and $m_i/m_e=400$
in (c). Consistent with simulations in smaller domains
\cite{Shay98b,Hesse99}, the rate of reconnection is insensitive to the
electron mass.

\begin{figure}
  \epsfig{file=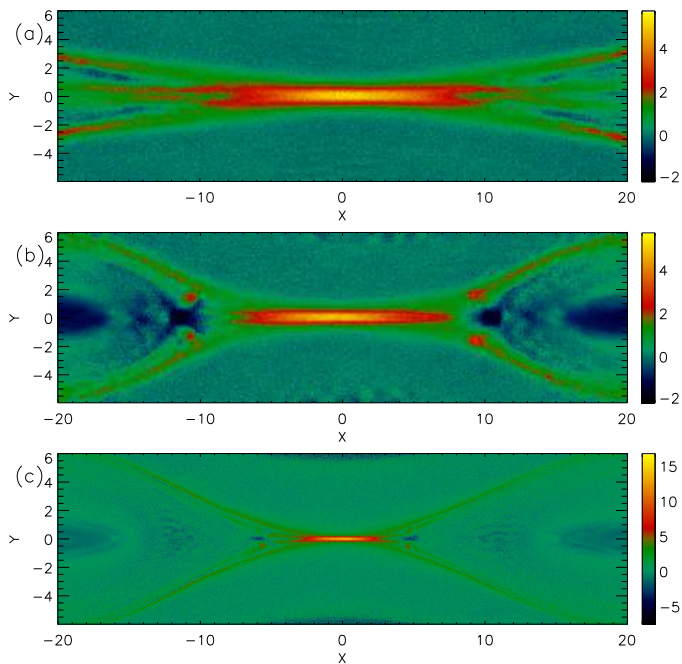,width=9cm}
  \caption{(color online). Blowups around the x-line of the
    out-of-plane electron velocity for: (a) $m_i/m_e=25,$ simulation
    size $204.8 \times 102.4,$ (b) $m_i/m_e=25,$ $51.2 \times 25.6,$
    and (c) $m_i/m_e=400,$ $51.2 \times 25.6.$}
  \label{jez}
\end{figure}

We now proceed to explore the structure of the electron current
layer. Shown in Fig.~\ref{jez} is a blow-up around the x-line of the
out-of-plane electron velocity for $m_i/m_e=25$ and two simulation
domains, $204.8\times 102.4$ in (a) and $51.2\times 25.6$ in (b), and
for $m_i/m_e=400$ in a simulation domain of $51.2\times 25.6$ in
(c). All of the data is taken in the phase where the reconnection rate
and the lengths of the region of intense out-of-plane current are
stationary. Reconnection forms intense current layers that have a
well-defined length (half widths of around $7d_i$ and
independent of the size of computational domain for $m_i/m_e=25$) and
then open up forming the open outflow jet that characterizes Hall
reconnection \cite{Shay99,Birn01}. The current layer in the case of
$m_i/m_e=400$ in Fig.~\ref{jez}c is distinctly shorter than the
smaller mass ratio current layers in Fig.~\ref{jez}a,b, suggesting
that the length of the electron current layer depends on the electron
mass and would be shorter for realistic proton-electron mass ratios.

\begin{figure}
  \epsfig{file=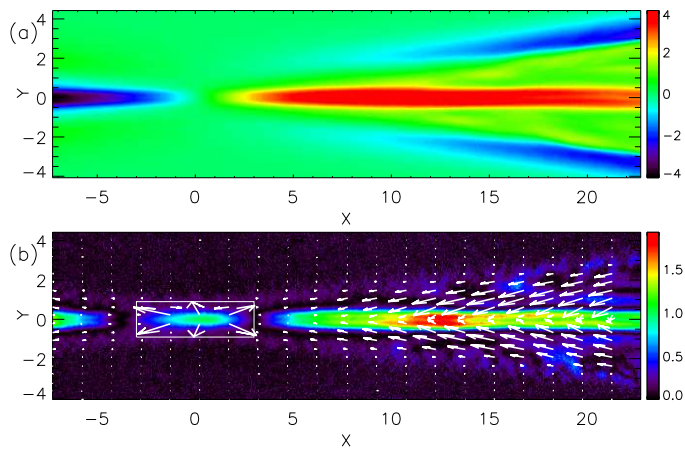,width=9cm}
  \caption{(color online). Blowups around the x-line for system size $204.8 \times
    102.4$ with $m_i/m_e=25.$ (a) The electron outflow velocity
    $v_{ex}.$ (b) Momentum flux vectors, ${\bf \Gamma} =
    p_{exz}\hat{\bf x} + p_{eyz}\hat{\bf y}$ (vectors in box
    surrounding x-line are multiplied by 20), with a background color
    plot of $\mid (E_z + ({\bf v}_e \times {\bf B}/c)_z)/E_z\mid.$}
  \label{vexp}
\end{figure}

Shown in Fig.~\ref{vexp}a is a blow-up around the x-line of the
electron outflow velocity $v_{ex}$ for the $m_i/m_e=25$, $204.8\times
102.4$ run corresponding to Fig.~\ref{jez}a. In contrast with the
out-of-plane current the electrons form an outflow jet that extends a
very large distance downstream from the x-line. This outflow jet
continued to grow in length until the end of the simulation. This
simulation, along with others at differing mass ratios, reveals that
the peak outflow velocity is very close to the electron Alfven speed
\cite{Shay01,Hoshino01b}.  One might expect that because of the
collimation of the outflow jet and its length, the reconnection rate
would drop. However, this is not the case. While there is an intense
jet in the core of the reconnection exhaust, the exhaust as a whole
quickly begins to open up downstream of the current layer ($J_z$). The
jet itself therefore does not act as a nozzle to limit the rate of
reconnection: the rate of reconnection remains constant even as the
length of the outflow jet varies in time.

\begin{figure}
  \epsfig{file=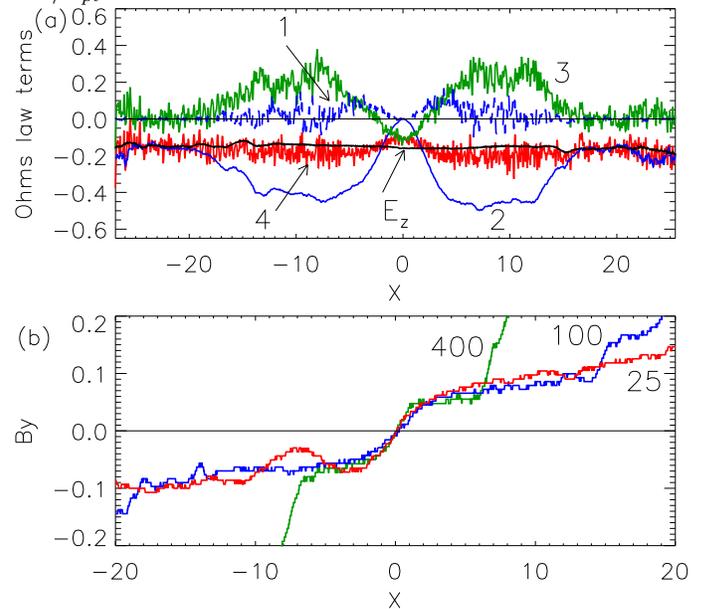,width=9cm}
  \caption{(color online). Results for simulation size $102.4 \times
    51.2$ with $m_i/m_e = 25$ and $100$; and $51.2 \times 25.6$ with
    $m_i/m_e = 400.$ (a) Cuts through the x-line of the contributions
    to Ohm's law for $m_i/m_e = 100.$ $1 \rightarrow -m_e/e {\bf v}_e
    \cdot \nabla v_{ez},$ $2 \rightarrow -\hat{\bf z} \cdot {\bf v}_e
    \times {\bf B}/e,$ $3 \rightarrow - \hat{\bf z} \cdot (\nabla
    \cdot {\bf P_e})/(n_e e),$ $ 4 \rightarrow $ sum of 1,2,3. (b)
    Cuts through x-line of $B_y$ for the three different $m_i/m_e$. }
  \label{ohms}
\end{figure}

To understand how the electrons can form such an extended outflow jet
while the out-of-plane current layer remains localized, we examine the
out-of-plane component of the fluid electron momentum
equation along the symmetry line of the outflow direction. In steady state
\begin{equation}
E_z=-\frac{m_ev_{ex}}{e}\frac{\partial v_{ez}}{\partial x}-\frac{1}{c}
v_{ex}B_y-\frac{1}{ne}{\bf \nabla}\cdot
{\bf\Gamma},
\label{emomentum}
\end{equation}
where ${\bf v_{e}}$ is the electron bulk velocity, ${\bf
  \Gamma}=p_{exz}\hat{\bf x}+p_{eyz}\hat{\bf y}$ is the flux of
z-directed electron momentum in the reconnection plane (not including
convection of momentum) with ${\bf p_e}$ the electron pressure tensor.
In Fig.~\ref{ohms}a we plot all of the terms in this equation along a
cut though the x-line along the outflow direction from a simulation
with $m_i/m_e=100$ and $L_x\times L_y=102.4 \times 51.2$. The data has
been averaged between $t=116.2$ and $t=117.0$. The electric field
(black) is balanced by the sum (red) of the electron inertia (dashed
blue), the Lorentz force (solid blue) and the divergence of the
momentum flux (green). The major contributions to momentum balance
come from the Lorentz force and the divergence of the momentum flux.
At the x-line the electric field drive is balanced by the momentum
transport \cite{Hesse99,Pritchett01}. The surprise is that the Lorentz
force, rather than simply increasing downstream from the x-line to
balance the reconnection electric field, instead strongly overshoots
the reconnection electric field far downstream of x-line. This
tendency was seen in earlier simulations \cite{Pritchett01} but there
was no clear separation of scales because of the small size of these
earlier simulations. Downsteam from the x-line the electrons are
streaming much faster than the magnetic field lines. Thus, in a
reference frame of the moving electrons the z-directed electric field
has reversed direction compared with the x-line. This electric field
tries to drive a current opposite to that at the x-line. Evidence for
this reversed current appears downstream of the x-line in
Fig.~\ref{jez}c. In spite of the strength of the effective electric
field, the reversed current carried by the electrons is small. As at
the x-line, the momentum transfer to electrons in this extended
outflow region is balanced by momentum transport. The momentum flux
around the x-line is shown as a 2-D vector plot in Fig.~\ref{vexp} for
the same run as in (a). The momentum flux has been multiplied by 20 in
the box surrounding the x-line. The data for this figure has been
averaged between $t=172.5$ and $174.5.$ The background color plot is
of $\mid (E_z + ({\bf v}_e \times {\bf B}/c)_z)/E_z\mid,$ which is $\gtrsim 1$
where the electrons are not frozen-in. Evident is the outward flow of
momentum around the x-line and the much stronger outward flow of
negative momentum in an extended downstream region. The momentum
transport is so large that the out-of-plane current downstream is
effectively ``blocked''. The force associated with this ``blocking
effect'' drives the flow of the large-scale jet of electrons
downstream of the x-line.

We define the length $\Delta_x$ of the inner dissipation region as the
distance from the x-line to the point where the Lorentz force
$v_{ex}B_y/c$ crosses the reconnection electric field $E_z$. At this
location the effective out-of-plane electric field seen by the
electrons reverses sign, causing the electron current $j_{ez}$ to be
driven in reverse, which allows the separatrices to open up. Thus, the
inner dissipation region defines the spatial extent of the magnetic
nozzle that develops during reconnection. Since the simulations
presented in this paper use artificial values of $m_e$, it is
essential to understand the $m_e$ scaling of $\Delta_x$ so that this
important length can be calculated for a proton-electron
plasma. The momentum equation of electrons in the outflow direction
yields a steady state equation for $v_{ex}$,
\begin{equation}
\frac{d}{dx}(\frac{1}{2}m_ev_{ex}^2)=\frac{e}{c}v_{ez}B_y,
\label{vexcalc}
\end{equation}
where $v_{ez}\sim c_{Ae}$. Thus, the profile of $B_y$ along the
outflow direction and its dependence on $m_e$ must be determined. This
profile is shown for $m_i/m_e=25$ (system size $102.4 \times 51.2$),
$100$ ($102.4 \times 51.2$) and $400$ ($51.2 \times 25.6$) in
Fig.~\ref{ohms}b.  Surprisingly, the profile of $B_y$ is apparently
independent of $m_e$.  Our original expectation was because of the
continuity of the flow of magnetic flux into and out of the x-line
that $B_y\sim B_0v_{in}/c_{Ae}\propto m_e^{1/2}$, where the outflow
velocity eventually rises to $c_{Ae}$. However, since the electrons
are not frozen into the magnetic field until far downstream, the
expected scaling fails. To calculate $v_{ex}$ we approximate $B_y$ by
a linear ramp and integrate Eq.~(\ref{vexcalc}). Setting the Lorentz
force equation to the reconnection electric field, we then obtain an
equation for $\Delta_x$,
\begin{equation}
\Delta_x=\left(\frac{m_e}{m_i}\right)^{3/8}\left(\frac{cE_z}{B_0c_A}\right)^{1/2}
\left(\frac{B_0}{d_iB_{y'}}\right)d_i.
\label{deltax}
\end{equation}
For the three simulations shown in
Fig.~\ref{ohms}b the simulations yield $2.9d_i$, $1.8d_i$ and $1.0d_i$
for $m_i/m_e=25$, $100$ and $400$, respectively, which is in
reasonable accord with the scaling. Extrapolating to a mass-ratio of
$1836$, we predict $\Delta_x\sim 0.6d_i$. In contrast the outer dissipation
region can extend to $10$'s of $d_i$.

We have shown that the electron current layer that forms during
reconnection stabilizes at a finite length, independent of the
periodicity of the simulation domain, and aside from transients from
initial conditions remains largely stable to secondary island
formation. Reconnection remains fast with normalized reconnection
rates of around $0.14$. The length of the electron current layer
$\Delta_x$ scales as $m_e^{3/8}$. Since the width $\delta$ of the
current layer scales with the electron skin depth $c/\omega_{pe}$, the
aspect-ratio $\delta/\Delta_x\propto (m_e/m_i)^{1/8}$. Extrapolating
from our $m_i/m_e=400$ simulations to $m_i/m_e=1836$ should not
significantly change the aspect-ratio and we therefore expect the
current layer to remain stable for real mass ratios.

The structure of the current layer is important to the design of
NASA's magnetospheric multiscale mission (MMS), which will be the
first mission with the time resolution to measure the electron current
layers that develop during reconnection. The length of the
out-of-plane electron current layer projects to around $c/\omega_{pi}$ for
a proton-electron plasma while the the outflow jet, which supports a
strong Hall (out-of-plane) magnetic field, extends $10$'s of
$c/\omega_{pi}$ from the x-line.

This work was supported in part by NSF, NASA and DOE.


{\bf Acknowledgments} This work was supported in part by NASA and the
NSF. Computations were carried out at the National Energy Research
Scientific Computing Center.


\end{document}